\documentstyle[11pt,newpasp,twoside]{article}
\def\edcomment#1{\iffalse\marginpar{\raggedright\sl#1\/}\else\relax\fi}
\reversemarginpar

\begin{document}
\title{Study on Neutron-induced Background in the CRESST Experiment}
\author{H. Wulandari, F. von Feilitzsch, M.
Huber, Th. Jagemann,  J. Jochum, T. Lachenmaier, J.-C. Lanfranchi,
W. Potzel, W. Rau, M. Stark, S. Waller} \affil{Physik Department
E-15, Technische Universit\"{a}t M\"{u}nchen, James-Franckstr.,
D-85747 Garching, Germany }

\begin{abstract}
CRESST II is an experiment for direct WIMP search, using cryogenic
detectors. The ratio of the two signals (temperature rise and
scintillation light) measured for each interaction is an excellent
parameter for discrimination of radioactive background. The main
remaining background is the neutron flux present at the
experimental site, since neutrons produce the same signals as
WIMPs do. Based on Monte Carlo simulations the present work shows
how neutrons from different origins affect CRESST and which
measures have to be taken into account to reach the goal
sensitivity.
\end{abstract}

\section{Sources of Neutron Background}
The flux of neutrons in the vicinity of CRESST (Jagemann et al.
2003) is dominated by low energy neutrons induced by radioactivity
in the surrounding rock/concrete. In addition, a very small
remaining impurity in the material used in the setup can induce
significant background if higher sensitivity is to be aimed for.

Although the flux of high energy neutrons induced by muons in the
rock is small compared to the total flux at the experimental site,
this component can penetrate a neutron moderator, reach the setup
and produce additional neutrons through spallations. Moreover,
neutrons can be produced by muons in the experimental setup,
particularly in the lead shield.

To investigate the contribution of each neutron source to the
expected background rate in CRESST II, a study based on Monte
Carlo simulations has been performed. For this purpose different
Monte Carlo codes have been employed, namely MCNP4B (Briesmeister
1997), MCNPX (Waters 1999), FLUKA (Fasso 2001) and MUSIC
(Antonioli 1999).

\section{Neutron Background Rates Expected in CRESST II}
The contributions of different neutron sources to the count rate
in the 15-25\,keV interval in a CaWO$_{4}$ detector are shown in
Table 1. In CRESST II, 60 GeV WIMPs with a cross section as
claimed by the DAMA experiment in (Bernabei et al. 2000) would
give 55 cts/kg/y in the same energy range. This makes it difficult
for CRESST II to check the DAMA evidence without a neutron
moderator. The flux of low energy neutrons from the surrounding
rock/concrete can be reduced effectively by a hydrogen-rich
material like polyethylene. For the CRESST II setup a polyethylene
shield (35-50 cm thick) is advisable. This will reduce the
background count rate in the CaWO$_{4}$ detector by up to three
orders of magnitude. Then the background will be dominated by
neutrons from other origins (see Table 1) and the sensitivity of
the experiment for the WIMP-nucleon cross section would be limited
to about $10^{-7}\,$pb.

\begin{table}      
\caption{Contributions of different neutron sources to the count
rate at 15-25\,keV in a CaWO$_{4}$ detector inside the CRESST
setup.} {\tabcolsep5pt
\begin{tabular}{l|c}
\tableline
Neutron origin& Count Rate \\
 & (cts/kg/y) \\
\tableline
Low energy neutrons from the rock: (i)\,\,no moderator &50\\
\hspace*{176pt}(ii)\,Pb/Cu + 50\,cm PE &0.04 \\
Low energy neutrons from fission of 0.1\,ppb $^{238}$U in the lead &0.2 \\
High energy neutrons produced by muons in the rock & 0.3  \\
High energy neutrons produced by muons in the shields & 1 \\
\tableline \tableline
\end{tabular}
\label{tabel1}}
\end{table}

The remaining neutron flux with the neutron moderator installed is
dominated by neutrons induced by muons in the lead shield. It is
possible to suppress this neutron background by detecting the
muons and rejecting muon coincident events. For CRESST II a muon
veto system is planned in addition to a neutron moderator. This
will enable CRESST II to reach the projected sensitivity of below
$10^{-7}\,$pb. The muon veto will be placed inside the
polyethylene shield and will have an efficiency of more than 90\%.

However, the muon veto will reduce the neutron background only by
a factor of three, unless high energy neutrons from the rock can
be overcome. Neutrons scattered in more than one detector can be
rejected as background, because WIMPs do not scatter multiply.
Therefore the neutron background will be further reduced and
multiple scattering can also be engaged to determine the remaining
single scatter neutron background. Such a simulation with an array
of detectors will be done in the near future to investigate, which
sensitivity level can be reached by this technique.

\end{document}